\documentclass[a4paper]{article}

\usepackage{18lomcon}        
\usepackage{cite}             
\usepackage{epsfig}           
\usepackage{epstopdf}

\newcommand{\be}{\begin{eqnarray}}
\newcommand{\ee}{\end{eqnarray}}

\begin{document}


\title{PRIMORDIAL BLACK HOLES AND COSMOLOGICAL PROBLEMS}

\author{Alexander Dolgov \email{dolgov@fe.infn.it}
}

\affiliation{ Novosibirsk State University, Novosibirsk, 630090,  Russia\\
ITEP, Moscow, 117218 Russia}


\date{}
\maketitle


\begin{abstract}

It is argued that the bulk of black holes (BH) in the  universe are primordial (PBH).  This assertion is strongly 
supported  by the recent astronomical observations, which allow to conclude that supermassive BHs  with
$M= (10^6 - 10^9) M_\odot$ "work" as seeds for galaxy formation,  intermediate mass BHs, 
$ M = (10^3 - 10^4)   M_\odot$, do the same job for globular clusters and dwarf galaxies, while black 
holes of a few solar masses are the constituents of dark matter of the universe.  The mechanism of PBH 
formation, suggested in 1993, which predicted such features of the universe, is described. The model leads to 
the log-normal mass spectrum of PBHs, which is determined by three constant
parameters. With proper adjustment of these parameters the above mentioned features are quantitatively 
explained. In particular, the calculated density of numerous superheavy BHs 
in the young universe, $ z = 5 - 10$,  nicely fits the data. The  puzzling properties of the sources of the
LIGO-discovered gravitational waves are also naturally explained assuming that these sources are PBHs.

\end{abstract}

\section{Introduction} \label{s-inro}

Recent, and not only recent, astronomical observations revealed many mysterious features of the 
universe, which were not expected in frameworks of conventional cosmology and astrophysics. 
All these problems are neatly solved if practically all black holes (BH) in the universe are 
primordial ones 
with a wide spread mass spectrum. Primordial black holes by definition are those which were formed
in the universe at prestellar epoch, i.e. before stars appeared in sky. The mechanism of their 
formation was suggested by Zeldovich and Novikov (ZN)~\cite{ZN}. According to them, PBH was formed
if the density fluctuation in the early universe was of order unity,  ${\delta\rho/\rho \sim 1}$, 
at the cosmological horizon scale. In this case the piece of space happened to be inside its 
gravitational radius, so it decoupled from the overall Hubble expansion and a black hole appeared.
In the original version PBH created by such mechanism were rather light (a small fraction of the solar
mass) and had narrow (delta-function) mass spectrum. At least such form of the spectrum was mostly 
assumed in subsequent analysis of observational manifestation of such PBHs.

In 1993~\cite{AD-JS} a generalization of ZN mechanism was proposed (see also later work~\cite{DKK}), which could lead to very massive 
PBH with log-normal mass spectrum:
\be
\frac{dN}{dM} = \mu^2 \exp{[-\gamma \ln^2 (M/M_0)], }
\label{dn-dM}
\ee
with only 3 parameters: ${\mu}$, ${\gamma}$,  ${M_0}$.
The form of the spectrum is practically universal. It is completely determined by the exponential cosmological expansion
during inflationary stage. 

Fitting the parameters of distribution (\ref{dn-dM}) one can explain the
accumulated astronomical data about black holes in contemporary and
young universe. In this sense the surprising results of the new precise 
observations performed in the recent decade are {\it predicted}
in the papers~\cite{AD-JS,DKK}.

In this talk I briefly review the following observational data.\\
{\it Young universe}, $z \approx 5-10$, overpopulated by:\\
{1. Bright quasi stellar objects (QSO), super-massive BHs.}\\
{2. Superluminous young galaxies.}\\
{3. Supernovae and gamma-bursters.} \\
{4. Very high level of dust.}\\
{ \it Contemporary universe:}\\
{1. Supermassive BH (SMBH) in every large galaxy.} \\
{2. SMBH in small galaxies and in almost {\it empty} space.} \\
{3. Stars older than the Galaxy and even older than the Universe. }\\
{4. MACHOs (low luminosity solar mass objects).} \\
{ 5. Problems with the  BH mass spectrum in the Galaxy: unexpected maximum at ${ M \sim 8 M_\odot}$. } \\
{6. Problems with the sources of the observed gravitational waves (GW). }\\
{7. Intermediate mass, ${ \geq 10^3 M_\odot}$, BHs in globular clusters and dwarf galaxies.}

More details and references can be found in~\cite{ad1,ad2}.

\section{Young universe}\label{s-young}
 
 \subsection{SMBH} \label{ss-SMBH}
About 40 quasars with ${z> 6}$ are already known, each quasar containing 
SBH with ${M \sim 10^9 M_\odot}$. 
The maximum redshift value among these quasars reaches
{${ z = 7.085}$}  i.e. the quasar was  formed 
before the universe reached {${0.75}$ Gyr. Its luminosity and mass are
respectively {${L= 6.3 \cdot 10^{13} L_\odot}$} and
${M=2 \cdot 10^9 M_\odot}$~\cite{qso-708}.} 
Other high $z$ quasars have similar properties.
The formation of SMBHs, which fuel these quasars
through the standard accretion mechanism, demands much more time
than the universe age at $z \sim 6$.
The unsolvable problem with creation of these SMBHs was 
multiply deepened with the discovery of a real "monster" 
of 12 billions solar masses~\cite{qso-630}, i.e. an order of 
magnitude more massive, than the mentioned above forty.
Even the formation of the contemporary SMBH, which had in their
disposal the whole universe age, 14 Gyr, is difficult to explain, see the 
next section.

After this Conference was over, a new discovery of a SMBH at now new
maximum redshift $z\approx 7.5$ and the mass 0.8 billion solar masses
was announced~\cite{qso-750}. This is not the largest mass in the
family of high redshift quasars, but what makes it particularly interesting is
that the surrounding matter is neutral, not ionized. This is a very strong
argument against formation of this SMBH by the usual accretion process,
so its primordial origin remains the only natural possibility.

 \subsection{Early bright galaxies} \label{ss-early-gal}
 
Several galaxies have been observed at high redshifts,
with natural gravitational lens ``telescopes''. A few examples are:\\
1) A galaxy at {${z \approx 9.6}$} which was created when the universe
was younger than 0.5 Gyr~\cite{gal-960}. \\
2) {A galaxy at {${z \approx 11}$}~\cite{gal-11} 
which already existed when the universe age was 
${t_U \sim 0.4}$ Gyr. It is particularly impressive that this very young galaxy is
three times more luminous in UV than other galaxies at ${z = 6-8}$.} 
This is a striking  example of unexpectedly early burn and powefull creature.\\
3) Not so young but extremely luminous galaxy was found three years ago. Its luminosity 
reaches gigantic magnitude, {{${L= 3\cdot 10^{14} L_\odot }$. The universe age when the galaxy
already existed was ${t_U \sim 1.3 }$ Gyr.} According to the authors of the 
discovery:
``The new study outlines three reasons why the black holes in the extremely luminous infrared galaxies,
 could have grown so massive. 
First, they may have been born big. In other words, 
{{the galactic seeds, or embryonic black holes, might be bigger than thought possible.''}
One of the authors, P. Eisenhardt said: ``How do you get an elephant?  
One way is start with a baby elephant."
{The BH was already billions of ${M_\odot}$ , when our universe was only a 
tenth of its present age of 13.8 billion years.}
``Another way to grow this big is to have gone on a sustained binge, consuming food faster than typically thought possible."  
For the realization of these conditions 
{low spin is necessary!}

According to the paper "Monsters in the Dark"~\cite{gal-monsters} 
{density of galaxies at ${z \approx 11}$ is 
${10^{-6} }$ Mpc${^{-3}}$, an order of magnitude higher than estimated 
from the data at lower z.}
{Origin of these galaxies is unclear.}

These data strongly support the idea that initially
primordial SMBHs  appeared and later galaxies were 
seeded by these PBHs. To the best of my knowledge this idea was 
first pronounced in ref.~\cite{AD-JS} and the recent observations 
do confirm the early creation of very massive black holes. 
 
 \subsection{Early miscellanea} \label{ss-misc}
 
 The universe at $z = 5 - 10$ was filled with
 supernovae, gamma-bursters, and was very dusty.
{To make dust a long succession of events is necessary:} 
{first, supernovae  exploded to deliver 
heavy elements into space (metals),} 
{then metals cool and form molecules,}
{and lastly molecules make macroscopic
pieces of matter.}
{Abundant dust is observed in several early  galaxies, e.g. in HFLS3 at 
${ z=6.34} $~\cite{dust-1} and in A1689-zD1~\cite{dust-2} at ${ z = 7.55}$.} 
The second galaxy is the earliest one where interstellar medium is observed
The universe age at this redshift is below 0.5 Gyr.

Catalogue of the observed dusty sources~\cite{dust-catalogue} 
indicates that their number is an order of magnitude larger 
than predicted by the canonical theory of galaxy evolution.

{Hence, prior to or simultaneously with the QSO formation a rapid star formation should take place.}
{These stars should evolve to a large number of
supernovae enriching interstellar space by metals through their explosions}
{which later make molecules and dust.}
{(We all are dust from SN explosions, but probably at much later time.)}

{Observations of high redshift gamma ray bursters (GBR) also indicate 
a high abundance of supernova at large redshifts.} 
{The highest redshift of the observed GBR is 9.4 and there are a few more
GBRs with smaller but still high redshifts.}
{The necessary star formation rate for explanation of these early
GBRs is at odds with the canonical star formation theory.}

\section{Mysteries in the sky today and in the nearest past} \label{s-today}

\subsection{Supermassive black holes}\label{ss-smbh}

{ Every large galaxy and some smaller ones 
contain a central supermassive BH with
mass typically  larger than} 
{ ${ 10^{9}M_\odot}$} in giant elliptical
and compact lenticular galaxies,
and {${\sim10^6 M_\odot}$} in spiral galaxies like Milky Way.
{The origin of these BHs is unclear.}
The accepted faith is that
these BHs are created by the matter accretion to galactic center with 
an excessive mass density.
However, the usual accretion efficiency is insufficient to create them during the Universe life-time, $t_U \approx 14$ Gyr. 
Even more puzzling is that  SMHBs  are observed  in small galaxies
 and even in almost empty space, where no material to make a SMBH can be found.

Below several examples are presented demonstrating serious inconsistencies
between observation and theoretical picture.

The mass of BH is typically 0.1\% of the mass of the stellar bulge of galaxy} 
but some galaxies may  have huge BH: {e.g. NGC 1277  has
the central BH  of  ${1.7 \times 10^{10} M_\odot}$, or ${60}$\% of its 
bulge mass~\cite{NGC1277}.
This creates serious problems for the
standard scenario of formation of central supermassive BHs by accretion of matter in the central part of a galaxy.

According to ref.~\cite{khan}
the galaxies, 
 Henize 2-10, NGC 4889,
and NGC1277 are examples of SMBHs at least an order of magnitude 
more massive than their host galaxy suggests. 
{The dynamical effects of such ultramassive central black holes are unclear. }

A recent discovery~\cite{strader} of an ultra-compact dwarf galaxy
older than 10 Gyr, enriched with metals, and probably with a massive 
black { hole} in its center also seems to be at odds with the standard model.


In the paper entitled
"An evolutionary missing link?  A modest-mass early-type
galaxy hosting an over-sized nuclear black hole"~\cite{loon},
a black hole with the mass ${M_{BH} = (3.5 \pm 0.8) \cdot 10^8 M_\odot}$,}
is found inside the host galaxy with mass of the stars
${M_{stars} = 2.5^{+2.5}_{-1.2} \cdot 10^{10} M_\odot}$,
and huge accretion luminosity: 
{${ L_{AGN} = (5.3 \pm 0.4) \cdot 10^{45} }$erg/s ${\approx 10^{12} L_\odot}$,}
equal to 12\% of the Eddington luminosity. 
The active galactic nuclei (AGN) is more prominent than expected for a host galaxy
of this modest size. The data are in tension with the accepted picture in which
this galaxy would recently have transformed from a star-forming disc galaxy into
an early-type, passively evolving galaxy.

Probably the most impressive in this list is a discovery of
``A Nearly Naked Supermassive Black Hole"~\cite{naked}.
According to the paper, a  compact symmetric radio source B3 1715+425
is too bright (brightness temperature ${\sim 3\times10^{10}}$ K at observing frequency 7.6 GHz) and too luminous (1.4 GHz luminosity 
${\sim 10^{25}}$ W/Hz) }
to be powered by anything but a SMBH, but its host galaxy is much smaller. 

There are more example of  such puzzling galaxies with superheavy black 
holes but even with the presented ones the inverted picture 
of galaxy formation looks more plausible, when first a 
supermassive black hole was formed and later it
attracted matter serving as a seed for subsequent galaxy formation.

\subsection{MACHOs} \label{ss-macho}

MACHO is the name of some invisible or low luminosity objects discovered 
through gravitational microlensing by Macho~\cite{MACHO2000,Bennet2005} 
and Eros~\cite{EROS2007} groups respectively in the Galactic 
halo and in the direction to the center of the Galaxy. Later they were registered
in the Andromeda (M31) galaxy~\cite{AGAPE2008}. The masses 
of the registered objects are about one 
half of the solar mass. 
The up to date situation with MACHOs is summarized in ref.~\cite{BDK}:\\
Macho group:
{ ${ 0.08<f<0.50 }$ (95\% CL)} 
{for ${0.15M_\odot < M < 0.9M_\odot} $;}\\
EROS: {${f<0.2}$, ${ 0.15M_\odot < M < 0.9M_\odot}$;}\\
EROS2: {${ f<0.1}$,~${10^{-6}M_\odot<M<M_\odot}$;}\\
AGAPE: {${0.2<f<0.9}$, 
for ${0.15M_\odot < M < 0.9M_\odot} $;}\\
EROS-2 and OGLE: {$ {f <0.1} $ for  ${M\sim 10^{-2} M_\odot}$ and
$ {f <0.2} $ for  ${ \sim 0.5 M_\odot}$.}

Thus, the  MACHO 
density is comparable to the density of the halo dark matter but their nature is unknown. They could be brown dwarfs,  dead stars, or primordial black holes. 
The first two options are in conflict with the accepted theory of stellar evolution, 
if such invisible stars were created in the conventional way.

The only remaining option is that MACHOs are low mass black holes,
but one can hardly imagine that such low mass black holes, 
abundant in the Galactic halo, were created as a result
of stellar collapse of normal stars. So the natural conclusion is that
MACHOs are primordial black holes as it is stated in ref.~\cite{DKK}.
The log-normal spectrum of the PBH allows to make much larger 
contribution to DM from heavier PBH, to which the microlensing 
method is not sensitive. So ultimately 100\% of DM may be made
out of PBHs with different masses.

\subsection{Properties of the sources of gravitational waves}\label{ss-GW}

Direct registration of gravitational waves (GW) by LIGO~\cite{GW-LIGO} 
revealed intriguing properties of the GW sources~\cite{BDPP}. 
The shape of the signal 
in the interferometer is well described by the assumption that the observed 
GWs are produced by the binary of coalescing BHs, but:
\begin{enumerate}{
\item{}
The origin of heavy BHs  with masses ${\sim 30 M_\odot}$ is unclear.
Such BHs are believed
to be created by massive star collapse, though a convincing theory is still lacking. To form so heavy BHs, the progenitors should have 
${M > 100 M_\odot}$ and  a low metal abundance to avoid too much
mass loss during the evolution. Such heavy stars might be present in
young star-forming galaxies but they are not yet observed in sufficiently high number.

\item{}
In all events, but one, the spins of the coalescing BHs are very small.
compatible with zero.
It strongly constrains astrophysical BH formation from close binary systems. 
However, the dynamical formation of double massive low-spin BHs in dense 
stellar clusters is not excluded, but difficult.

\item{}
Formation of BH binaries from the original stellar binaries has  very low
probability. Stellar binaries were 
formed from common interstellar gas clouds and are quite frequent in galaxies.
{If BH is created through stellar collapse,} {a small non-sphericity of the collapse results in a 
huge velocity of the BH and the binary is destroyed.}
{BH formation from PopIII stars and subsequent formation of BH
binaries with 
${(36+29) M_\odot}$ is analyzed and 
found to be negligible. } 

}\end{enumerate}

{All these problems are solved if the observed sources of GWs are the binaries of primordial black holes (PBH).}

\subsection{Globular clusters and intermediate mass BHs.}\label{ss-glob-clust}

Recently the so called intermediate mass black holes (IMBH) with masses 
$M \approx 2000 M_\odot $  and $M \sim 20000 M_\odot $ were presumably 
observed in the centers of globular clusters~\cite{BH-sh-s,bh-gc-2}. 
These observations nicely fit our conjecture~\cite{AD-KP-gc} 
that IMBH play an important role in the formation and evolution of  globular 
clusters. Using the parameters of the mass distribution (\ref{dn-dM}), found
in our paper~\cite{BDPP}, we find that the density of the primordial IMBH
is sufficient to seed the formation of all globular clusters observed in galaxies.

In addition to globular clusters, IMBHs are probably also contained in 
centers of dark stellar
clusters~\cite{central-imbh-1,central-imbh-2}. These clusters have high
mass-luminosity ratio. They may be the remnants of dwarf spheroids
with the masses between those of  globular clusters and large
galaxies. It looks natural that these spheroids were seeded by the
primordial IMBH~\cite{AD-KP-gc}.

\subsection{Solar mass Black holes in the Milky Way} \label{ss-bh-gal}

The mass spectrum of black holes observed in the Galaxy demonstrates
some peculiar features, which are difficult to explain in the standard model of BH
formation by stellar collapse. In particular, it is found~\cite{BH-gal1} 
that the masses of black holes in the Galaxy are concentrated in the narrow
range ${ (7.8 \pm 1.2) M_\odot }$.
This result agrees with another paper where
{a peak around ${8M_\odot}$, a paucity of sources with masses below
 ${5M_\odot}$, and a sharp drop-off above
${10M_\odot}$ are observed}~\cite{BH-gal2}. 

On the other hand, such mass spectrum is well described by the log-normal
form. This is an argument in favor of primordial origin of the
black holes in the Galaxy.

\subsection{Old stars in the Milky Way}\label{ss-old-stars}

Recently several groups presented substantially more accurate determinations
of stellar ages in the Galaxy. Surprisingly quite a few stars happened to be
considerably older than expected.

According to  ref.~\cite{star-1}:
employing thorium and uranium  abundances
in comparison with each other and with several stable elements the age of
metal-poor, halo star BD+17$^o$ 3248 was estimated as
${13.8\pm 4}$ Gyr. This star is much older than the inner halo of the
Galaxy, which has the age equal to ${11.4\pm 0.7}$ Gyr~\cite{age-halo}. 

The age of another star in the galactic halo, HE 1523-0901, was estimated 
to be about 13.2 Gyr~\cite{star-2}.
First time many different chronometers, such as the $U/Th$, 
$U/Ir$, $Th/Eu$, 
and $Th/Os$ ratios to measure the star age, have been employed.

And at last a star older than the universe  was found~\cite{star-3}.
Metal deficient {high velocity} subgiant in the solar neighborhood
HD 140283  has the age {${14.46 \pm 0.31 }$ Gyr.}
{The central value exceeds the universe age by two standard deviations,
if ${H= 67.3}$, and ${t_U =13.8}$ Gyr;}
while if ${H= 74}$, and ${ t_U = 12.5} $ Gyr, the star would be older than the
universe by more than 10 ${\sigma}$. This is of course impossible, but
the star may look older that it is, if initially the star was enriched by 
heavy elements and evolve to its present state faster than the normal one.
Our model of PBH formation~\cite{AD-JS,DKK} leads also to creation of compact primordial
stellar-like objects consisting not only from hydrogen and helium but enriched 
with plenty of heavier elements.

 \section{Mechanism of massive PBH formation} \label{s-MPBH-form}

In this Section the main features of the mechanism~\cite{AD-JS,DKK} 
of massive PBH formation are described. We assume that a slightly modified baryogenesis 
scenario suggested by Affleck and Dine (AD)~\cite{BG-AD} 
is realized. The main ingredient of this AD-scenario is a scalar field $\chi$
with non-zero baryonic number $B$. It is assumed that the potential of $\chi$
has the so called flat directions along which the potential does not rise. In the course of the
cosmological expansion $\chi$ might acquire large expectation value
turning practically into a classical field with large $B$. Later after decay of
$\chi$ this accumulated  baryonic number turned into baryonic number of  
quarks, leading to a large baryon asymmetry $\beta$ of the universe. 
It may be even
of order unity, while the observed value of $\beta$ is about $10^{-9}$,

We modified the AD mechanism by introduction of general renormalizable 
coupling of $\chi$ to the inflaton field $\Phi$ (the first term in the r.h.s.
of the equation below), which can be written in the form:
\be 
U = {g|\chi|^2 (\Phi -\Phi_1)^2}  +
\lambda |\chi|^4 \,\ln( \frac{|\chi|^2 }{\sigma^2 })
+\lambda_1 \left(\chi^4 + h.c.\right) + 
(m^2 \chi^2 + h.c.).
\label{U-AD}
\ee
With this interaction the flat direction of the potential $U$ are open only 
when $\Phi \approx \Phi_1$, which was taken by $\Phi$ in the course of 
inflation before it was over.
If the window to flat direction, when ${\Phi \approx \Phi_1}$ is open only 
{during a short period,} cosmologically small but possibly astronomically large 
bubbles with high ${\beta}$ could be created, These bubbles with large $\beta$
might occupy only a small fraction of the universe volume,
while the rest of the universe would have the
normal small baryon-to-photon ratio
{${{ \beta \approx 6\cdot 10^{-10}}}$, created 
by small ${\chi}$}, which did not succeed to penetrate through the briefly open  window to 
a large value.

After the QCD phase transition, when massless quarks turned into heavy 
nucleons, the initial isocurvature perturbation created by inhomogeneities 
in the chemical content turned into (large) density perturbations.
This would lead to an early formation of PBH or compact 
stellar-type objects with high baryonic density. As a result,
the bulk of baryons and 
maybe antibaryons would be contained in compact cosmologically tiny stellar-like objects or PBH.
These high-B density bubbles would live in huge by size but not so dense universe with low baryonic background
density, which initially was practically homogenenous.

The formation of PBHs or compact stellar type objects
took place at very high ${z}$ after the QCD phase transition at 
${T \sim 100}$ MeV down to ${T \sim  }$ keV.

As a byproduct, the mechanism of refs.~\cite{AD-JS,DKK}
may lead, though not necessarily, to abundant compact 
antimatter objects in the universe and, in particular, in the Galaxy~\cite{BDK,anti-1,anti-2}.

\section{Conclusion}\label{concl}

The problems emerged from the multitude of astronomical observations, some of which
are mentioned in this talk, are uniquely and simply resolved if the universe is
populated by the primordial massive black holes and stellar-like
compact objects with wide mass spectrum. The mechanism which leads to an
abundance of such objects in the universe was put forward  in 
1993~\cite{AD-JS,DKK} and essentially predicted the subsequent 
surprising discoveries.

All the multitude of the various astronomical data are well explained by the 
natural baryogenesis model which leads to formation of PBHs 
and compact stellar-like objects in the early universe after the QCD phase 
transition, ${t \leq 10^{-5}} $ sec.
These objects are predicted to have log-normal mass spectrum.
They can be numerous enough to give significant contribution to the
cosmological dark matter or even make all of it.

The model opens the possibility for
the inverted picture of the galaxy formation, when
firstly supermassive black holes are formed which later accrete  matter 
creating galaxies. The new observations persuasively indicate in this direction.
Lighter PBHs with 2000 ${M_\odot}$ are predicted in sufficient amount to 
explain the origin of globular clusters, while heavier PBHs, with 
$ M \sim 10^4 M_\odot$ can seed formation of dwarf spheroids. There seem
to be strong indications in favor of this scenario.

{PBHs formed through such mechanism can explain the peculiar features of the 
sources of GWs observed by LIGO.}

The considered mechanism resolves the numerous mysteries of ${z \sim 10}$ 
 universe: abundant population of supermassive black holes, 
early created gamma-bursters and supernovae, early bright galaxies, and evolved chemistry including dust.

Existence of high density invisible "stars" (MACHOs) is explained.

"Older than ${t_U}$" stars may exist. The old age is mimicked by the unusual initial chemistry. 

The model can possibly lead to the prediction of numerous  compact antimatter 
objects (antistars). The observational data  allow for large amount of such objects in the Galaxy. However, their density is model dependent and the prediction is uncertain.

\section*{Acknowledgments}

This work was supported by RSF Grant No. 16-12-10037.


\end{document}